\documentclass[twocolumn,showpacs,preprintnumbers,amsmath,amssymb]{revtex4}

\usepackage{graphicx}
\usepackage{dcolumn}
\usepackage{bm}

\begin{document}

\preprint{APS/123-QED}

\title{Measurement of the ortho-positronium confinement energy in mesoporous thin films}

\author{Paolo  Crivelli}
 \email{paolo.crivelli@cern.ch}
\affiliation{%
Instituto de Fisica, Universidade Federal do Rio de Janeiro (UFRJ),
Rio de Janeiro, 21941-972, Brazil\\
}%

\author{Ulisse Gendotti}
 \email{ulisse.gendotti@cern.ch}
\author{Andr\'e Rubbia}
\affiliation{Institut f\"ur Teilchenphysik, ETHZ, CH-8093 Z\"urich, Switzerland
\\
}%
\author{Laszlo Liszkay}
\altaffiliation{On leave from KFKI Research
Institute for Nuclear and Particle Physics, P.O. Box 49, H-1525
Budapest, Hungary}
\author{Patrice Perez}
\affiliation{CEA, Saclay, IRFU, 91191 Gif-sur-Yvette Cedex, France
\\
}%
\author{Catherine Corbel}
\affiliation{LSI, Ecole Polytechnique, Route de Saclay, 91128, Palaiseau Cedex, France
\\
}%

\date{\today}

\begin{abstract}
  In this paper, we present measurements of the ortho-positronium emission energy in vacuum from mesoporous films using the time of flight technique. We show evidence of quantum mechanical confinement in the mesopores that defines the minimal energy of the emitted Ps. Two samples with different effective pore sizes, measured with positron annihilation lifetime spectroscopy, are compared for the data collected in the temperature range 50-400 K. The sample with smaller pore size exhibits a higher minimal energy ($73\pm$5 meV), compared to the sample with bigger pores ($48\pm$5 meV), due to the stronger confinement.
The dependence of the emission energy with the temperature of the target is modeled as ortho-positronium being confined in rectangular boxes in thermodynamic equilibrium with the sample. 
We also measured that the yield of positronium emitted in vacuum is not affected by the temperature of the target.

\end{abstract}

\pacs{Valid PACS appear here}
\maketitle

\section{Introduction}

Positronium (Ps), the bound state of electron and positron, was extensively investigated since its discovery \cite{deutsch} contributing to the development of bound state QED (see e.g. \cite{savely} for a review on the current status of this field). Furthermore, this system provided stringent limits on possible deviation from the Standard Model that could indicate new physics \cite{oPsInv} (see also \cite{andre} for a comprehensive review of former experiments).
Moreover, in the field of materials science Ps found various applications due to its unique properties (see e.g. \cite{barbiellini,gidley1} for modern reviews on this subject). One recent example is the characterization via positron annihilation lifetime spectroscopy (PALS) of low-k dielectrics that are potential candidates for the next generation of integrated circuits \cite{gidley2,gidley3}.
 
The motivation of the work presented in this paper is to understand if mesoporous silica with an interconnected pore network could be used for producing a high fraction of positronium at low temperatures. This would open the door for a new generation of experiments in fundamental research. Cold positronium could be used to improve the precision of spectroscopic studies of Ps \cite{chu,fee} or to perform the first spectroscopy of the Ps$_2$ molecule \cite{cassidy}.
Furthermore, it could provide an alternative to the methods that are used for anti-hydrogen formation \cite{ATHENA}-\cite{ALPHA}.
 As it was suggested sometime ago \cite{humberston}-\cite{deutch}, anti-hydrogen could be formed using charge-exchange of Ps with anti-protons. This was demonstrated for the charge conjugate reaction \cite{merrison} and in the ATRAP collaboration resonant charge exchange collisions of positrons with Rydberg Cs atoms were used to form Rydberg Ps that via charge exchange with the anti-protons produced anti-hydrogen in Rydberg states \cite{Storry:2004zz}.
Recently, two experiments \cite{AEGIS,AIP} were proposed to perform an anti-gravity test using this process. In both experiments, one of the main issues is that a high fraction of Ps at low temperatures should be available.
Another interesting application would be the possibility to perform an experiment in order to confirm the interpretation of the recent DAMA/LIBRA annual modulation signal \cite{DAMALIBRA,foot} as generated by Mirror type dark matter \cite{opsDM}. A step further would be to achieve Bose-Einstein condensation of Ps \cite{mills1}. This would allow the exploration for the first time of the effects of the collective properties of a matter-antimatter system.

In mesoporous silica, Ps is produced by injecting positrons into the film and the distribution of the implantation depth follows a Makhovian profile \cite{nieminen}. In the following we solely consider the long lived triplet spin state (called ortho-positronium with 142 ns lifetime, o-Ps) because the singlet spin state (called para-positronium, p-Ps) has a very short lifetime of 125 ps and can be considered as annihilating in the target. The o-Ps (for simplicity we will refer to it as Ps) that diffuses into the pores loses its kinetic energy via scattering. If the pores are interconnected, the Ps has a probability to tunnel from one pore to another. A fraction of the Ps reaches the film surface and exits into vacuum. 
 A classical model of thermalization process was developed by Nagashima et al. \cite{nagashima1}. Their calculations reproduce very well the behavior for SiO$_2$ aerogel with pore sizes of about 100 nm. However, a classical approach is not expected to give reliable predictions for Ps confined in few nm pores because quantum mechanical effects become relevant. As a matter of fact, in this regime the de Broglie thermal wavelength of Ps is comparable with the size of the pores. Recently, Mariazzi et al. \cite{mariazzi} considered phonon scattering to reproduce the thermalization process of Ps in a box (closed porosity) showing that the minimal energy is not that of the lowest accessible level because the momentum phonons can exchange is fixed. In the same paper, they pointed out that this is not the case in rectangular channels because in one direction the side of the potential well tends to infinity (z-axis). Therefore, the magnitudes of the $k_x$ and $k_y$ momentum are quantized but the $k_z$ tends to be a continuum and the minimal energy that Ps can reach is given by the ground state in the x-y components. Thus measuring the Ps emission energy provides a method to distinguish between the two different pores architectures (see Section \ref{sec:discussion}).        

In previous studies of Ps emission in vacuum using time of flight (TOF),
many interesting effects, such as the emission from the surface of different
materials \cite{tuomisaari}-\cite{mills2} were investigated. Recently, this technique was applied to study mesoporous and hybrid silica films \cite{yu}-\cite{xu2}, to evaluate the continuity barrier \cite{tanaka} and the effect of thermalization for pore surfaces decorated with different groups \cite{he2}. However, the influence of the temperature on Ps emission in vacuum from mesoporous thin films was never studied in detail. To our knowledge, only a work of Mills et al. \cite{mills2} with fumed silica revealed a dependence with the temperature of the Ps emission in vacuum. The fraction of Ps in the low energy tail was estimated but no quantitative estimate on the value of the minimal emission energy was obtained. Very recently Cassidy et al. \cite{cassidyPhysRevA} measured the emission energy of Ps from mesoporous films using Doppler spectroscopy. This is a different technique from that used in the present study. Results consistent with the one presented in this paper (though not sample temperature dependencies) have been obtained.

\section{Description of the experimental apparatus}\label{sec:setup}

\subsection{Positronium production}
In this paper, we study the Ps yield and emission energy in vacuum as a function of the target temperature for two different kinds of mesoporous thin films with the same tetraethoxysilane (TEOS) mineral source for the silica network skeleton precursor: CTACl-TEOS and F127-TEOS. The density of the C sample is approximately 1.2 g/cm$^3$ and of the F sample is 1.5 g/cm$^3$. Both samples were spin-coated on glass similar to the ones we measured in \cite{APL1}. The C samples are prepared via a sol-gel process using cetyl trimethyl ammonium chloride (CTACl) cationic surfactants as the organic pore generator (porogen) agent \cite{cohen}. A pure aqueous method is used. The CTACl/TEOS molar ratio for the films prepared is 0.22. After deposition, the CTACl-TEOS/Glass samples are treated at 130 $^0$C and stored in air. The F samples use non-ionic Pluronic F-127 triblock copolymer (EO$_{106}$PO$_{70}$EO$_{106}$) as surfactant and were prepared in the same way as described in \cite{F127}.
Both samples were calcinated for 15 minutes at 450 $^0$C in air immediately before the e$^+$ measurements.  The recorded X-ray diffraction patterns indicate no symmetry in the pore organization.

\subsection{Slow positron beam}
\begin{figure}[h!]
\begin{center}
\hspace{.0cm}\includegraphics[width=.5\textwidth]{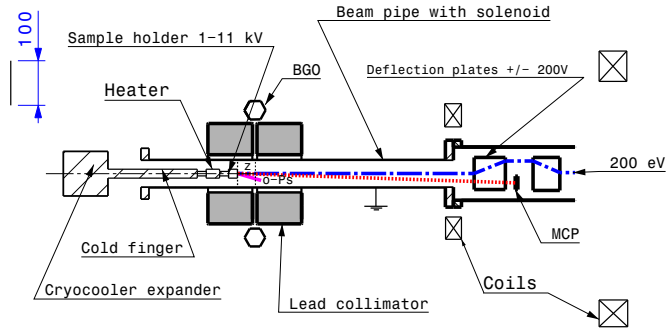}
\caption{\em (Color online) Experimental setup for the TOF measurements (the scale is in mm). The dashed line is the trajectory of the incoming positrons (blue) and the dotted line is the one of the secondary electrons (red).}
\label{traj}
 \end{center}
\end{figure}

The ETHZ slow positron beam used for these measurements is described in greater detail in \cite{beamNIMA}. The positrons flux is 25000 e$^+$/s. The slow positron beam is stopped in the SiO$_2$ target. The positrons can either form positronium, i.e. o-Ps or p-Ps, or annihilate into $2\gamma$'s. 
The detection with an MCP (Micro-channel Plate) of the secondary electrons (SE) emitted when the positrons hit the target serves for tagging the positronium formation. The SE leave the target accelerated to 1-11 keV by the same voltage applied to the target relative to the grounded transport tube that is used to implant the positrons in the positronium converter. The SE are then transported by a magnetic field in the backward direction, as shown in Figure \ref{traj}. The electrons move along the magnetic field line in spirals and are deflected to the MCP region by the $E\times B$ filter. The tagging efficiency  varies from 70\% to 30\% in the energy range 1-10 keV.

The samples are mounted on a cryocooler head to allow the possibility of varying their temperature in the range of 50-400 K. 

\subsection{PALS and TOF detectors}

The start time t$_0$ for the detectors is triggered by the MCP detecting the SE emitted when the positrons hit the target. In both PALS
and TOF detectors the stop is given by one (or more) annihilation photons
depositing some energy in the calorimeter (ECAL). Both ECAL are composed
of BGO crystals with hexagonal shape, 61 mm external diameter and 200 mm
length. The time resolution of the system MCP-ECAL was measured to be
around 5 ns (FWHM). The typical energy resolution of the crystals is about
25-30$\%$ (FWHM). 
The ECAL for the TOF is placed behind a lead slit at a
distance z from the target that can be varied (as sketched in Figure \ref{traj}). The detector is screened by four half cylinders of lead surrounding the beam pipe. The thickness of the shielding is 70 mm, the width of the slit is set to 5 mm and its position with respect to the target can be adjusted. For
this measurement in order to maximize the signal to background ratio the center of the slit was placed at a distance of 18 mm from
the target. The main contribution to the background is given by photons coming from direct positrons and p-Ps annihilations in the target (so called prompt peak). Some of those photons can be detected in the crystals after Compton scattering in the lead shield. The determination of the slit position with respect to the target was done by scanning the slit position in 0.1 mm steps and recording the maximum of the 511 keV annihilation peak. The scans were performed 30 min after the temperature was set on the sample. This technique was very important in order to correct the slit position as a function of the temperature. In fact, at 50 K the contraction of the cryocooler head was measured to be $1.30\pm 0.05$ mm in
agreement with the prediction of finite elements calculations performed with the COMSOL package \cite{comsol}. 
 The PALS detector was designed to have a large acceptance to provide a uniform efficiency for detecting the Ps emitted in vacuum \cite{cugPaper}.

\section{Monte Carlo simulations}\label{sec:MC}
In order to design the detectors and interpret the data, simulations served as a powerful tool.
In the Monte Carlo (MC) simulations of our setup, the 3D EB-fields were calculated with the COMSOL multi-physics program and the positron/electron trajectories in the beam were simulated with GEANT4. The simulation of the photon detection in the apparatus was based on the same package \cite{geant}. New classes were written in order to simulate the o-Ps production, propagation in the beam pipe, reflection on pipe walls that were assumed to be Knudsen-like (Ps is reflected isotropically), pick-off effect or decay (Ps is not a standard particle in GEANT4). The events for the o-Ps$\to 3 \gamma$ process were generated taking into account the decay matrix element \cite{ORE}. The geometries of the beam transport pipe, photon detector, positron tagging system and its material were coded into simulations. The results were cross-checked with our experimental measurements for both photon detection \cite{oPsInv,cugPaper,3body,pc} and particle transport in the EB-fields \cite{beamNIMA}. 
 
\section{Results}
The measurements were taken in a clean vacuum of $10^{-9}$ mbar. To avoid water contamination of the film during the cooling down, the target was kept at room temperature for an hour using a heater before lowering its temperature. The cooling cycles were repeated and the data confirmed the reproducibility of the results. The signals from the BGO's photomultipliers are split to record both energy and timing with a charge-to-digital converter (QDC CAEN v792) and a time-to-digital converter (TDC CAEN v775). A cut on the energy deposited in the BGO between $300<E_{BGO}<550$ keV was applied to optimize the signal to background ratio suppressing Compton scattering events in the collimator from direct and p-Ps annihilations in the target. 

\subsection{PALS measurements}
\begin{figure}[!htb]
\begin{center}
\hspace{.0cm}\includegraphics[width=0.5\textwidth]{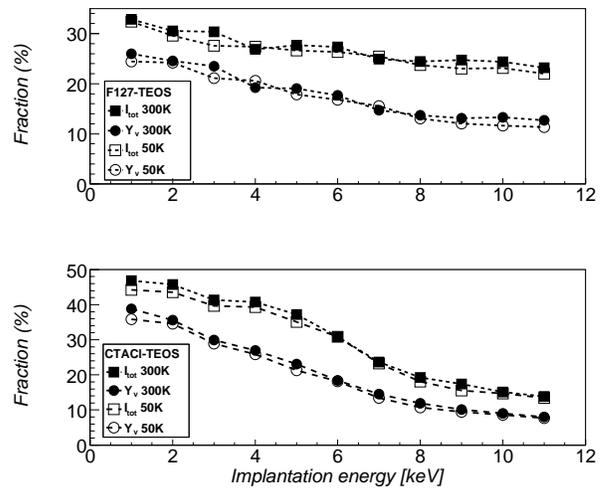}
\vspace*{-.cm}
\caption{\em Yield of Ps emitted in vacuum ($Y_v$) and total yield of Ps ($I_{tot}=I_2+I_3$) for F (upper plot) and C samples (lower plot) at room temperature and 50 K.}
\label{PALSdata}
 \end{center}
\end{figure}

In this section we present the results we obtained from the PALS measurements.
The spectra are analyzed  using the LT9 \cite{lt9} program. 
 The lifetimes of the decay components and its fractions are resolved by fitting the PALS spectra. The program finds 3 exponentials convoluted with a 5 ns FWHM resolution function of the spectrometer. The shortest exponential (less than 4 ns) is originated by direct positron and p-Ps annihilations. It is disregarded because we are interested only in contribution of o-Ps. 
 We define as ($\tau_2$, $I_2$) ,and  ($\tau_v$, $I_3$) the intensities and the lifetimes of the two longer exponentials. 
To determine the yield $Y_v$ of Ps emitted in vacuum and the lifetime  $\tau_f$ in the pores of the film we used a model of Ps escaping in vacuum \cite{APL}. 
According to this model the lifetime $\tau_f$ in the pores of the film is defined as:
\begin{equation}
\tau_f = {\big(({\tau_2}^{-1}-{\tau_v}^{-1})I_2/(I_2+I_3)+{\tau_v}^{-1}\big)}^{-1}
\end{equation}
The yield $Y_v$ of Ps emitted in vacuum  was calculated according to:
\begin{equation}
Y_v = (I_2+I_3) \kappa_v /({\tau_f}^{-1}+\kappa_v)
\end{equation}
where 
\begin{equation}
\kappa_v=({\tau_2}^{-1}-{\tau_v}^{-1})I_3/(I_2+I_3)
\end{equation}
 is the escape rate of free Ps into vacuum.
We extracted the films thicknesses for the two samples by fitting the total Ps yield as a function of the positron implantation energy to the Makhovian profile. These thicknesses are reported in Table \ref{tbl1} in which we also present the values of $\tau_f$ and  $\kappa_v$  calculated with the expressions above. We use the results of the fits of the PALS spectra at 6 keV for the C sample and 10 keV for the F sample at 50 K and 300 K. These implantation energies will serve as a reference for the rest of the paper. We choose those values in order to maximize the amount of thermalized Ps. At these energies the majority of the positrons are still implanted within the films (no significant drop of the total Ps yield $I_{tot}$ is observed, see Fig. \ref{PALSdata}) and the emission energy is in the constant region (see next Section).
With $\tau_f$ one can calculate the effective pore size $a$ in the films applying the Gidley et al. \cite{gidley2,gidley4} extension of the Tau-Eldrup model \cite{tau}-\cite{ito} (we will call it hereafter RTE model).
As one can see, at lower temperature the lifetime in the film increases as predicted by the RTE model. This can be understood in the following way: the overlap of the Ps wave function with the volume contained within a distance $\delta=0.18$ nm  \cite{gidley2,gidley4} from the walls for which the annihilation rate is assumed to increase is less for Ps confined in the pores occupying the ground state than for Ps in excited states. Since at lower temperatures the population of the ground state is higher the pick-off rate decreases. However, the measured lifetimes are lower than what is expected by the calculations using the RTE model. For the C sample the discrepancy is less than 10\% but for the F sample the measured value is 30\%  lower suggesting that the pick-off rate increases. Deviations from the RTE model were already observed in previous measurements using the 2 to 3$\gamma$ ratio technique \cite{mariazzi2,fischer} and in a measurement using PALS \cite{thraenert}. Different factors could be responsible for that as pointed out in those papers and deserves further studies.

 Since the parameter $\delta$ of the RTE model was calibrated at room temperature we use the lifetimes at 300 K to determine the effective pore size that Ps experiences. We report the results at the end of the paper in Table \ref{tbl3} for both rectangular channels and cubic boxes.
Interestingly, we found that the yield of Ps emitted in vacuum can be considered as independent of the temperature of the target (see Fig. \ref{PALSdata}). 
This may seem to contradict the fact that the lifetime in the films increases with the temperature. However, our measurements indicate that the escape rate in vacuum is smaller at lower temperature (see Table \ref{tbl1}), explaining the observation that the yield is constant. 
A possible explanation of this effect can be found considering that Ps tunnels from one pore to the other. In this case, the tunneling probability will decrease with temperature explaining our measurements. 

\begin{table}
\caption{\label{tbl1} Film thickness (Z), Ps lifetime in the pores $\tau_{f}$ and escape rate $\kappa_v$ at 50 K and 300 K.}
\begin{ruledtabular}
\begin{tabular}{cccccc}
 Sample & Z[nm] & $\tau_{f}^{300K}$[ns] & $\tau_{f}^{50K}$[ns] & $\kappa_{v}^{300K}$[$\mu$s$^{-1}$] & $\kappa_{v}^{50K}$[$\mu$s$^{-1}$]\\
\hline
C &  700$\pm$200 & 54$\pm$1 & 60$\pm$1 & 27$\pm$1 & 25$\pm$1 \\
F &  1000$\pm$200 & 74$\pm$1 & 82$\pm$1 & 17$\pm$1 & 13$\pm$1 \\
\end{tabular} 
\end{ruledtabular}
\end{table}

\subsection{TOF measurements}

\begin{figure}[h!]
\begin{center}
\hspace{.0cm}\includegraphics[width=0.5\textwidth]{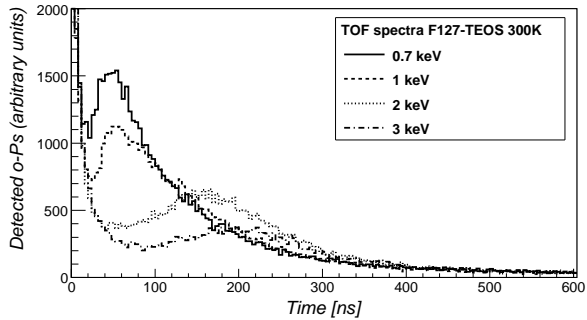}\caption{\em TOF spectra of the F sample for positron implantation energies of 0.7-1-4-10 keV.}
\label{TOF_Fd}
 \end{center}
\end{figure}

\begin{figure}[h!]
\begin{center}
\hspace{.0cm}\includegraphics[width=0.5\textwidth]{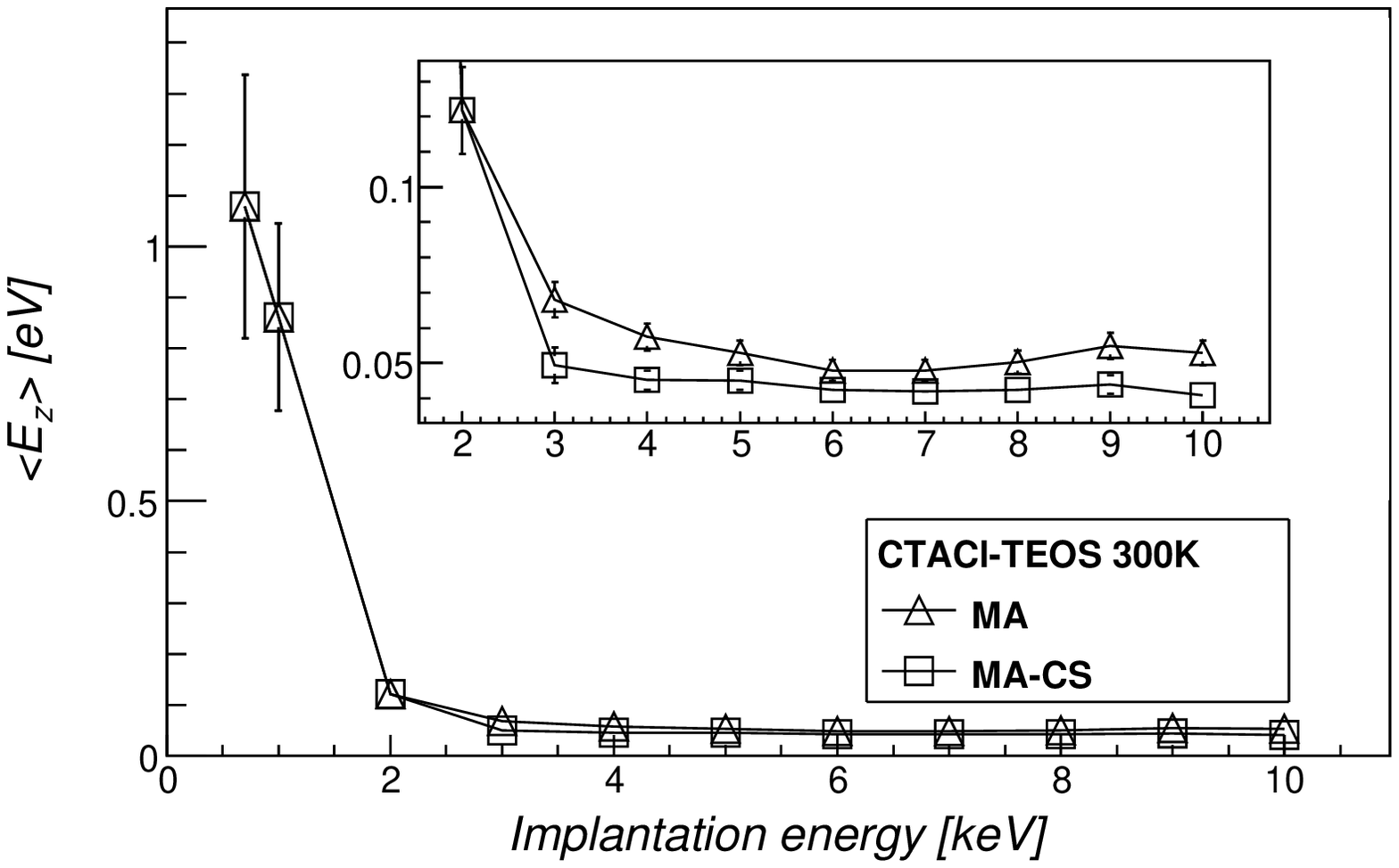}
\hspace{.0cm}\includegraphics[width=0.5\textwidth]{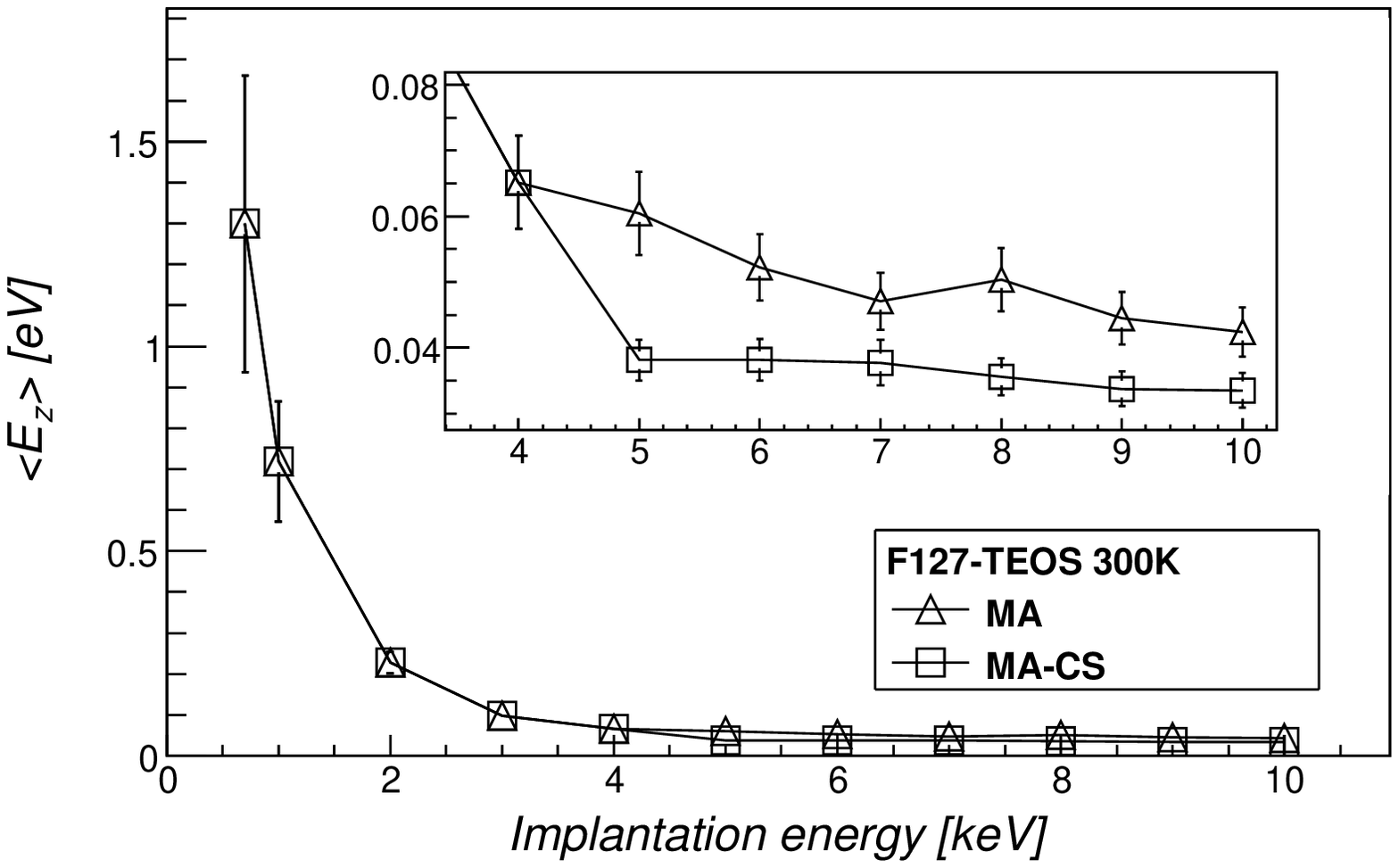}
\caption{\em  Positronium mean emission energy  $<E_z>$ as a function of the implantation voltage at a target temperature of 300 K. The triangles represents the energy extracted with the maximum analysis method (MA), the squares are after the subtraction of the non-thermalized part (maximum analysis method after correction of the spectra, MA-CS).}
\label{oPsEvsV}
 \end{center}
\end{figure}

\begin{figure}[h!]
\begin{center}
\hspace{.0cm}\includegraphics[width=0.5\textwidth]{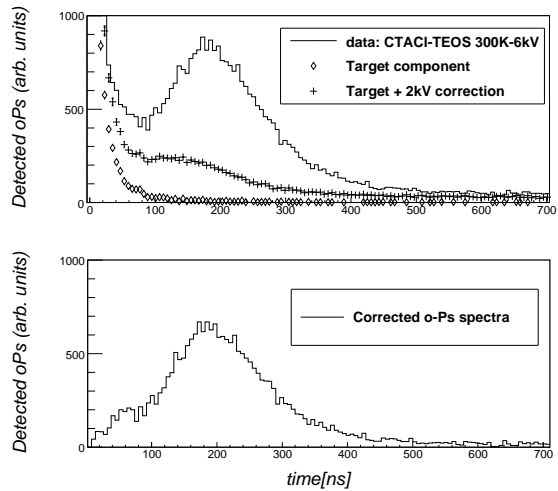}
\caption{\em Upper plot: the solid line is the TOF spectra without correction, the diamonds is the contribution from the target and the crosses is the sum of the contribution from the target and the non-thermalized part. Lower plot: the TOF spectra after the correction of the target and the non-thermalized components.}
\label{components}
 \end{center}
\end{figure}

Figure \ref{TOF_Fd} shows the acquired TOF spectra at different implantation energies for the F sample. For every run the time spectra of each crystal are calibrated finding the position of the peak arising from the annihilations in the target that defines t$_0$.
First we extract the mean energy of Ps emission in vacuum applying the analysis method proposed in \cite{mills2} as follows. The background due to the annihilations in the target is subtracted by fitting the measured time spectra with the resolution function of our detector determined using targets (aluminum and kapton) in which the Ps formation is negligible. The TOF spectra are corrected for the Ps decays and the time spent in front of the detector with the factor $1/t \cdot e^{+t/ \tau _{v}}$. The maximum of the peak distribution defines the mean Ps emission energy in the direction perpendicular to the film surface. Let us note that, neglecting the reflection of Ps in the beam pipe, with the time-of-flight method one measures only the mean of the energy component perpendicular to the collimator that we define as the z-axis (we define it as $<E_z>$). The triangles in Fig. \ref{oPsEvsV}, represent it as a function of the implantation energy.  
One can see that starting from 3 keV for the C sample (this was confirmed by other measurements on similar samples \cite{cassidyPhysRevA}) and 4 keV for the F sample the value of the energy emitted in vacuum tends to be constant. Clearly, the emission energy of Ps calculated in this way is not the minimal energy due to the confinement in the pores because one has different contributions given by the convolution of the emission energy with the implantation profile. In order to isolate the thermalized part from the TOF spectra, one has to subtract the contributions of non-thermalized Ps components (see Fig. \ref{components}). To estimate these non-thermalized contributions, we suppose that the shape of their distribution, NT(t), is represented by the TOF spectrum obtained at a low implantation energy. To select the implantation energy (from the ones we had measured) and the scaling factor of NT(t) for the subtraction, we relied on the MC. We used the values for which the best fit between the MC and the spectra obtained after the subtraction of NT(t) was achieved (for more details see \cite{thesiscug}). We found that the best fits were obtained for 2 keV with L$_{2keV}$=200 nm in the case of the C sample, and 3 keV and L$_{3keV}$=350 nm for the F sample. For a given implantation energy E$_i$, NT(t) is scaled down by the fraction of positrons implanted at depths smaller than L$_{2keV}$ and L$_{3keV}$. This was determined by using a Makhovian profile.

The results of this analysis are shown as the squares in Fig. \ref{oPsEvsV}.
A parabolic fit is used to determine the position of the maximum. The statistical error of the fit is typically $\pm$9 ns for the F and $\pm$6 ns for the C sample. 
The uncertainty on the determination of the slit position of $\pm0.1$ mm results in a systematic error of the order of $\pm1$ meV in the determination of the Ps mean emission energy $<E_z>$ (for implantation energies above 3 keV). 
The subtraction procedure described above introduces a systematic error
 that we estimated analyzing the data using different values of L$_{2keV}\pm50$ nm and L$_{3keV}\pm50$ nm. The estimated error is $\pm2.2$ meV for the C and  $\pm1.6$ meV for the F sample.
Thus the combined statistical and systematic error is at a level of $\pm$2.9 meV for the F and $\pm$3.0 meV for the C sample. 

\begin{figure}[h!]
\begin{center}
\hspace{.0cm}\includegraphics[width=0.5\textwidth]{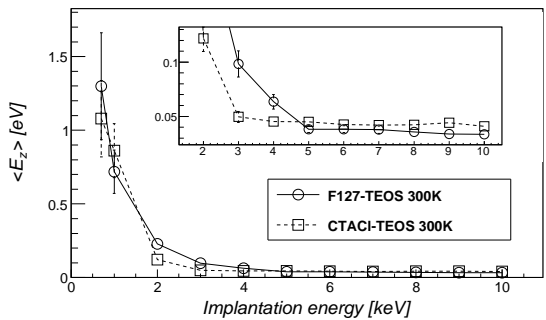}
\caption{\em Ps mean emission energy $<E_z>$ as a function of the positron implantation energy for C and F samples at 300 K.}
\label{oPsEvsV_RT}
 \end{center}
\end{figure}

\begin{figure}[h!]
\begin{center}
\hspace{.0cm}\includegraphics[width=0.5\textwidth]{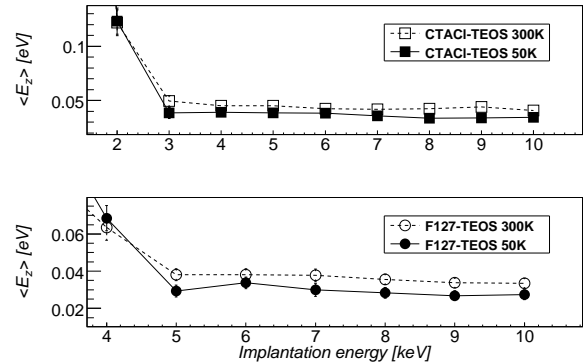}
\caption{\em Upper plot: Ps mean energy $<E_z>$ for implantation energies higher than 2 keV at 50 K and 300 K for the C sample. Lower plot: Ps mean energy as a function of the implantation energies higher than 4 keV at 50 K and 300 K for the F sample.}
\label{oPsEvsV_LT}
 \end{center}
\end{figure}

The results for the C and the F samples at room temperature and at 50 K are shown in Figs. \ref{oPsEvsV_RT}-\ref{oPsEvsV_LT}. For implantation energies higher than 4 keV for the C and 5 keV for the F sample the values $<E_z>$ of the mean emission energy are constant. In the C sample, $<E_z>$ reaches its constant value at lower implantation voltages because the pore size is smaller than in the F sample (see Fig. \ref{oPsEvsV_RT}). Those values are higher than the thermal energy that Ps will have if it would thermalize at the temperature of the film. As expected in the presence of confinement in the pores, the mean emission energy is higher for the C sample with pore sizes smaller than the F sample.    

\begin{figure}[h!]
\begin{center}
  \hspace{.0cm}\includegraphics[width=0.5\textwidth]{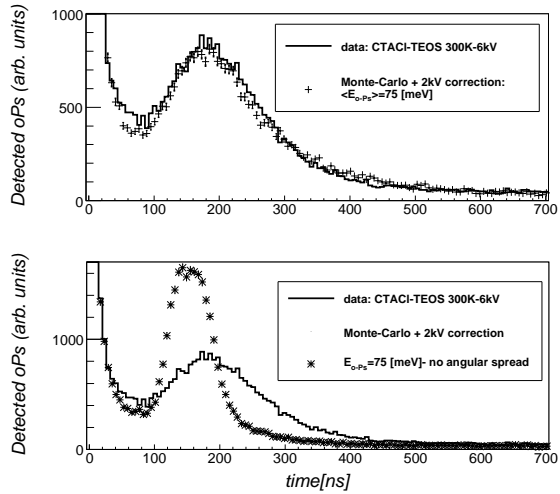}
\caption{\em Upper plot: comparison between the data of the C sample at 6 keV and the MC simulating mono-energetic Ps emitted isotropically from the film surface. Lower plot: comparison between the data of the C sample at 6 keV and the MC simulating mono-energetic Ps emitted perpendicular from the film surface. In both cases, the measured non thermalized part (called 2 keV correction in the legend) was added to the MC.
}
\label{MC}
\end{center}
\end{figure}

The TOF technique measures $<E_z>$, the mean energy of the Ps atoms in the z-direction. To find the mean emission energy of Ps in vacuum, $<E_z>$ should be multiplied by a factor $\xi$ that takes into account the angular distribution. Assuming that the Ps is emitted mono-energetically and isotropically from the surface, one can calculate that $\xi$ is equal 2. In this estimation, the reflection of Ps in the beam pipe and the detector acceptance, i.e. the fact that a fraction of events decaying before or after the collimator aperture are detected, are not taken into account. Therefore, to determine $\xi$ considering these effects we used the MC simulation we described in Section \ref{sec:MC}.  As shown in the upper plot of Fig. \ref{MC}, a satisfactory agreement between the data and the MC (adding the spectra at 2 keV that takes into account the non-thermalized Ps) is achieved. We attribute the difference between 40 and 100 ns to the approximation used in the subtraction method where the contribution of the non-thermalized Ps is underestimated since only a spectrum of a defined energy is used for this correction. The fact that the Ps is emitted with an angular spread is clearly supported by the data. As one can see in the lower plot of Fig. \ref{MC} for Ps emitted with no angular spread the data are not reproduced. The physical interpretation is that in the films we studied the pores have no organization thus they are expected to be randomly aligned. The value of $\xi$ estimated with the MC is 1.7. This is consistent with the expectation of the analytical result and the values reported in previous experiments \cite{nagashima2,yu,ito1,tanaka,cassidyPhysRevA}. 

A detailed scan shows that the mean Ps energy ($<E_{PS}>=\xi<E_z>$) decreases with the sample temperature down to a minimum level (see Fig. \ref{oPsEnergyvsT}). For the C sample this value is basically constant (73$\pm$5 meV) in the range of temperature in which we performed our measurement. This can be understood by the fact that in this sample the confinement energy is much higher than the thermal energy at room temperature ($kT\simeq 25$ meV) thus almost all Ps is in the ground state. For the F sample there is a weak dependence on the temperature. Due to the bigger pore size compared to the C sample, the energy of the ground state is only twice the thermal energy at room temperature. Therefore, the probability to find the Ps occupying an excited state is higher. As expected, this probability decreases with the temperature thus the minimal energy reaches its constant value of 48$\pm$5 meV. 

The time that Ps spends in the films before being emitted in vacuum was not considered in our determination of the emission energy. The measurements presented in Fig.  \ref{oPsEnergyvsT} are in a regime in which a classical approach is not expected to give reliable results. Some theoretical work to develop a full quantum mechanical picture of the emission process is required to address this problem (as pointed out in \cite{cassidyPhysRevA} as well).

\subsection{Discussion}\label{sec:discussion} 
To understand the behavior of the value of the minimal energy as a function of the film temperature we present a simple model of Ps in thermodynamic equilibrium at a temperature T in rectangular boxes. 

The expectation value $< H >$ of the Hamiltonian operator for Ps confined in 1 dimensional infinite well in contact with a reservoir at a temperature $T$ is given by:  
\begin{equation}
< H > = k{T^2} \frac{1}{Z}\frac{dZ}{dT} 
\end{equation}
where Z is the partition function defined as
\begin{equation}\label{eq:Z}
Z(a)= { }\sum _{n=1}^{\infty }{{e }^{- \frac{{h^2}{n^2}}{8 m {a^2}}/k T}}
\end{equation}
where $a$ is the dimension of the well, $m$ is the Ps mass, $n$ is the principal quantum number and $h$ and $k$ are the Planck and the Boltzmann constants. 
To calculate the mean value $<H>$ of the energy for the 3D case we can use
\begin{equation}
<H>= <{H_x}>+ <{H_y}>+ <{H_z}>
\end{equation}
where to calculate $<{H_y}>$ and $<{H_z}>$ one can substitute the pore side length $a$ in Eq. \ref{eq:Z} with $b$ and $c$.  
For Ps confined in rectangular pores, we thus obtain:
\begin{equation}\label{eq:H}
<H>=k{T^2}\Big(\frac{1}{Z(a)}\frac{dZ(a)}{dT}+ \frac{1}{Z(b)}\frac{dZ(b)}{dT}+ \frac{1}{Z(c)}\frac{dZ(c)}{dT}\Big)
\end{equation}
For the case of a cubic box one can set $<H_x>=<H_y>=<H_z>$.

\begin{figure}
\begin{center}
\includegraphics[width=0.5\textwidth]{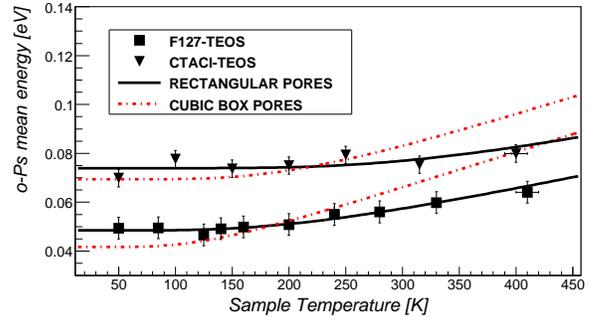}
\caption{\em (Color online) Positronium mean energy as a function of the mesoporous film temperature. Those results are obtained at 6 keV for the C and 10 keV for the F sample. The solid lines are the results of the using Eq. \ref{eq:H} with the pore side lengths $a$,$b$,$c$ left as free parameters. The dashed lines were obtained fitting with Eq. \ref{eq:H} with a single side length free $a=b=c$ (cubic box pores).}
\label{oPsEnergyvsT}
\end{center}
\end{figure}

To compare the prediction of the pore size that one can extract from the TOF measurements with the PALS results we fit the data using Eq. \ref{eq:H}. To construct the function used for the fit we kept only the first 50 terms of the sum. This is very conservative, the probability for Ps to occupy a state higher than n$>$10 for the kind of target and the temperatures we used in this study is already negligible.
The solid lines in Fig. \ref{oPsEnergyvsT} represent the fit to the data where the pore side length ($a$,$b$,$c$) are left as free parameters. The fits were repeated assuming cubic box pores and the results are shown as the dashed lines in Fig. \ref{oPsEnergyvsT}. We used MIGRAD from the MINUIT package \cite{minuit} as a minimization procedure.
As one can see, the fit to the data suggests that the pores of both samples are better modeled as rectangular pores than cubic boxes. As proposed in \cite{cassidyPhysRevA}, to compare the values obtained from the fit with the ones extracted from the PALS measurements, one has to add twice the parameter $\delta$ (see previous Section). The pore side lengths obtained in this way are reported in Table \ref{tbl3}.
In particular, the fit supports the idea that the pores are better modeled by rectangular channels. The pore side length in one direction ($c$) obtained from the fit is much longer than the substrate thickness. Since, the same values for the side length $a$ and $b$ are obtained fitting the data assuming rectangular channels instead of rectangular boxes we do not report this value in Table \ref{tbl3} as it has no physical meaning. 

\begin{table}
\caption{\label{tbl3} Comparison of the pore sizes obtained from the PALS measurements at 300 K and from the fit of the Ps mean emission as a function of the temperature of the sample (see Fig. \ref{oPsEnergyvsT}) for both cubic box (BOX) and rectangular pores (RECT). Minimal energy of Ps $<E_{PS}>$ for which the errors are the combined statistical and systematic error.}
\begin{ruledtabular}
\begin{tabular}{ccc}
\multicolumn{3}{c}{{\bf C Sample}}\\
 & TOF & PALS\\
$a_{BOX}$  [nm] & 3.3$\pm$0.1 & 4.2$\pm$0.5 \\
$(a,b)_{RECT}$ [nm] & (2.7$\pm$0.8, 2.7$\pm$0.8) &  (3.3$\pm$0.5, 3.3$\pm$0.5)\\
$<E_{Ps}>$ [meV] & 73$\pm$5 \\
\hline
\multicolumn{3}{c}{{\bf F Sample}}\\
 & TOF & PALS\\
$a_{BOX}$  [nm] & 4.1$\pm$0.1 & 6.4$\pm$0.5\\
$(a,b)_{RECT}$ [nm] &(2.9$\pm$0.8, 3.6$\pm$1.8) & (4.3$\pm$0.5, 5.3$\pm$0.5) \\
$<E_{Ps}>$ [meV] & 48$\pm$5\\
\end{tabular}
\end{ruledtabular}
\end{table}

According to the quantum mechanical model for Ps thermalization of Mariazzi et al. \cite{mariazzi} the index $n$ in the sum of Eq. \ref{eq:Z} could differ from 1 for cubic pores if the level separation of two close Ps energy levels is higher than the maximum momentum that a single phonon can exchange. We performed fits with different $n>1$ but the results did not improve, supporting the idea that the pores are not very well modeled by cubic boxes.    
To summarize, for the C sample with the TOF data the best fit was obtained for pores modeled as square channels of 2.7$\pm$0.8 nm side length while the lifetime method gives 3.3$\pm$0.5 nm. The best fit to the data for the F sample was obtained for rectangular channels with a cross section of 2.9$\pm$0.8 nm by 3.6$\pm$1.8 nm that has to be compared with 4.3$\pm$0.5 by 5.3$\pm$0.5 of the PALS measurement. This pore size was extracted by applying the RTE model to reproduce the measured lifetime of 74 ns assuming the ratio $a/b$ to be the same as in the TOF results. Both values obtained with the TOF measurement are systematically lower than the PALS results. Nevertheless, considering the approximation used in our model, the assumptions that the pores can be treated as rectangular channels and the uncertainty in the determination of the pore size we conclude that our results (summarized in Table \ref{tbl3}) are in reasonable agreement with expectations.

\section{Conclusions}
In this paper, we show that the yield of Ps emitted in vacuum measured with the PALS technique is independent of the temperature of the mesoporous thin films.
The lifetime in the films increases with a decrease in temperature but the measured values are lower than expected by the RTE model. This suggests that another source of pick-off which depends on the temperature should be invoked. Further studies are needed to investigate the origin of this effect. The escape rate in vacuum decreases explaining the observation that the yield of Ps emitted in vacuum is the same at 50 and 300 K. 

Furthermore, we show that due to quantum mechanical confinement in the pores the Ps emission energy into vacuum has a minimal value. The minimal energy is higher for the sample with smaller pores and this constant value is reached at a lower positron implantation energy. 
Our results are in fair agreement with a model of Ps confined in rectangular channels in thermal equilibrium with the sample. The measured minimal energy for the C sample was found to be 73$\pm$5 meV while for the F sample it is 48$\pm$5 meV. These experimental results provide a solid ground to understand how to produce Ps at lower temperature and could serve to develop a more realistic model to interpret the data.

\section*{Acknowledgments}
This work was supported by the Swiss National Foundation, the ETH Zurich and the CNPq (Brazil). P.C. would like to thank C. Lenz Cesar, P. Lotti and B. Barbiellini for the very useful discussions.

\end{document}